This manuscript has been accepted for publication in the International Journal of Energy, Environment, and Economics

# RESOURCE-OPTIMIZED AND ENERGY-AWARE AGENTIC AI FRAMEWORK ANCHORED ON BLOCKCHAIN FOR SECURE SOFTWARE SUPPLY CHAINS

***Toqeer Ali Syed***
***Asadullah Abdullah Khan****
Faculty of Computer and Information System,
Islamic University in Madinah, Kingdom of Saudi Arabia

## ABSTRACT

This paper proposes a blockchain-backed agentic security framework designed to safeguard the complete software development lifecycle (SDLC) while also securing the agentic AI components responsible for monitoring it. The framework coordinates a set of specialised security agents, covering source integrity, dependency and SBOM analysis, CI configura tion auditing, artifact verification, and runtime policy evaluation, each supported by a large language model (LLM) that interprets artefacts, reasons over tool outputs, and produces structured security reports. To ensure agent trustworthiness, every agent generates a cryptographically signed attestation that is recorded in a permissioned blockchain via smart contracts, including an agent registry, an immutable attestation log, and an enforceable release-policy module. Communication among agents and with blockchain nodes is secured using a consortium-operated certificate authority, ensuring authenticated and tamper-resistant interactions. A detailed use-case and sequence flow demonstrate how a source code security agent performs analysis, anchors its attestation on-chain, and triggers a verifiable allow/block deployment decision. The proposed framework of fers decentralised integrity transparent provenance, uninterrupted security assurance and a generalisable architecture to incorporate the agentic AI into the modern software supply chain security.

---

* Corresponding Author's E-mail: Asadullahkhanomarkhil@gmail.com

# INTRODUCTION

The software systems of the modern world rely on the complex and distributed ecosystem of source code repositories, third-party dependencies, build pipelines, artifact registries, and cloud-native deployment environments. This increased interdependence has made the attack surface to have enormous growth thus allowing adversaries to hack into software at any point in its lifecycle [1, 2, 3]. Cases like the SolarWinds breach showed that detection based methods are no longer enough to address hostile activities as attackers are now instead targeting the cycle of development instead of the system deployed [1]. The latest supply-chain models and frameworks, such as the NIST secured Software Development Framework (SSDF) [4] and the OpenSSF Scorecard project [5], formulates continuous verification, provenance, and auditing as staples of reliable software.

In parallel, the rise of agentic artificial intelligence has enabled autonomous and semi-autonomous systems capable of performing multi-step tasks, orchestrat ing tools, and analysing complex artefacts throughout the software development process [6, 7]. These agents are particularly well-suited for security workflows, where they can interpret repository changes, generate and analyse SBOMs, inspect CI configurations, validate artifact signatures, and infer misconfigurations using large language models (LLMs) [8]. However, the introduction of such agents also raises a new set of trust questions: How can organisations verify that agents acted correctly? How can the integrity of their security assessments be enforced? And how can the system detect compromised or malicious agents? Recent work on agentic AI security has shown that malicious instructions can also propagate through agent-to-agent messages and tool-mediated workflows, creating prompt-injection risks that require provenance-aware validation [36].

Researchers have come up with attestation-based trust models to counter these difficulties, which lock the software artefacts to cryptographically verifiable evidence of their creation and validation [9, 10]. Concurrently, blockchains and distributed records have become potent primitives to tamper-resistant provenance tracking, multi-party trust and decentralised policy enforcement [11, 12]. The combination of these technologies allows an ecosystem in which not just the artefacts of software but also the security agents that operate upon them can be verifiable entities and the operation of these verifiable entities can be fully traced.

This paper introduces a decentralised, blockchain-backed agentic framework for securing the entire software supply chain lifecycle. The proposed system deploys specialised security agents covering source integrity, dependency analysis, CI/CD configuration, artifact verification, and runtime policy checks each supported by an LLM capable of contextual reasoning across heterogeneous security data. Every agent produces a cryptographically signed attestation describing its findings, which is anchored in a permissioned blockchain. A release-policy smart contract evaluates these attestations to determine whether a given build should be trusted for deployment.

Our framework contributes a unified architecture that secures both (i) the software supply chain and (ii) the agentic AI ecosystem itself. Through decen tralised attestation, consensus-backed state, and CA-secured communication, agents become first-class verifiable entities rather than opaque components. A detailed use-case illustrates how a source-code security agent analyses repository changes, generates an LLM-supported risk assessment, anchors evidence on-chain, and participates in a verifiable release decision. This establishes a new foundation for trustworthy, autonomous security systems that operate continuously across the SDLC.

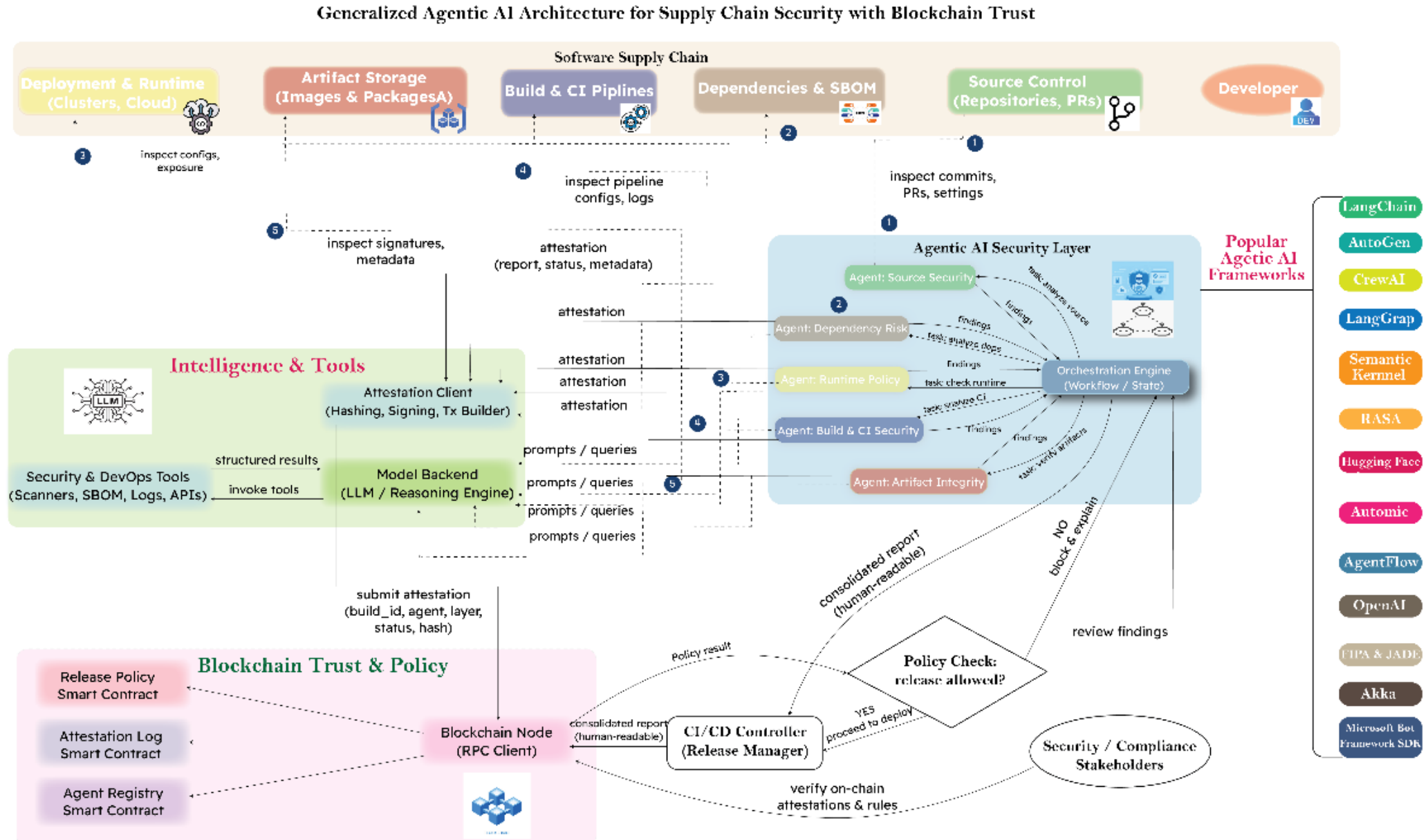

**Figure 1 Generalised agentic AI architecture for software supply chain security with blockchain-backed attestations. This figure provides an introduction-level overview of the SDLC, the agentic security layer, the LLM and tools, and the blockchain trust layer.**

# Background

## Software Supply Chain and Its Vulnerabilities

A software supply chain is a collection of entities, activities, and relationships that can be used in the development, construction, testing and packaging, and deployment of software systems depicted in figure 2. This consists of the source code repositories, dependency registries, continuous integration and continuous deployment (CI/CD) pipelines, artifact storage, and update distribution mecha nisms [13, 14]. All these provided elements are possible attack surfaces by the adversaries.

The weaknesses that are common to software supply chains are malicious dependency injection, hacked build server, misconfigured environments, or unauthorized code commits [15, 16]. Incidents like the SolarWinds backdoor and the XZ Utils intrusion exemplify the fact that attackers are going after the development phase and the build phase more and more than deployed code and break trusted components long before they become public [1, 17]. After being introduced into the pipeline, such attacks may spread downstream using signed, apparently valid artifacts, and therefore become very difficult to detect [9].

The traditional mitigation techniques like code examination, formal analysis, and artifact dynamics are only able to guarantee integrity and traceability, but they are mostly reactive in nature [18]. Though provenance verification is improved using frameworks such as SLSA and in-toto, they do not have the autonomy to prevent or react to dynamic attacks, such as injection attacks, insecure deserialization, weak access control, and misconfigured pipelines [19]. This constraint highlights the importance of adaptive, autonomous systems capable of continually learning and applying security in the process of software development [20].

## Agentic Artificial Intelligence

The term agentic AI refers to a system of autonomous agents that perceive their own environment, reason about contextual data and take actions against defined security targets [21, 22]. Unlike traditional automation scripts, agentic systems operate through closed feedback loops of perception, cognition, action and learning that permit dynamic response to changing environments [23]. Agents could be independent or co-operating and communicate either via protocols or knowledge repositories [24].

In the context of cybersecurity, self-directed and explainable defense is a possibility provided by agentic AI [20]. Agents can track repositories, analyse commit behaviour, ensure configuration integrity, and isolate anomalies in build pipelines. Through cognitive reasoning and self-reflection, these agents will be able to defend their actions, determine the level of risks, and modify defensive policies, depending on experience gained. This feature makes agentic AI especially fit to protect distributed and constantly changing software supply chains [21, 25].

## Reinforcement Learning for Autonomous Defense

Reinforcement learning (RL) constitutes the algorithmic foundation of em powerment of agentic systems [26]. In RL, an agent plays with an environ ment that is a Markov Decision Process (MDP), or which is defined by the pairM = (S,A,P,R,γ), where S denotes the set of states, A the set of possible actions, P(s′|s,a) the transition probabilities, R(s,a) the reward function, and γ the discount factor for future rewards. The purpose of the agent is to acquire an optimal policy π∗(a|s) that should maximize a cumulative reward [27]:

$$\pi^* = \underset{\pi}{argmax} E\left[\sum_{t=0}^{\infty} \gamma^t R(s_t, a_t)\right]$$

Applying RL to the framework of software supply chain security enables the agents to derive defensive strategy by constant interaction with the environments of CI/CD [28, 29]. As an example, an RL agent monitoring suspicious API calls or unusual repository behavior can alter access controls, isolate suspicious builds or initiate re-verification steps. Gradually, such agents come to have adaptive defense policies not only to decrease the chances of successful attacks but also to decrease the false positive [30].

## Integration of RL and Agentic Systems in Supply Chain Defense

Through a combination of agentic AI and its cognitive capabilities with the component of adaptive decision-making in RL, software supply chain security could transform into self-defensive software [22, 21]. Within this model, the agents do not only guarantee compliance like checking signatures or depen dencies, but also detect and address vulnerabilities, such as injection attacks, insecure deserialization, and misconfigurations in real-time. Multi-agent col laboration, common-ground knowledge representation, and reinforcement-based learning gives an opportunity to have scalable, explainable, and resilient defense mechanisms that can secure complex software ecosystems [20, 25].

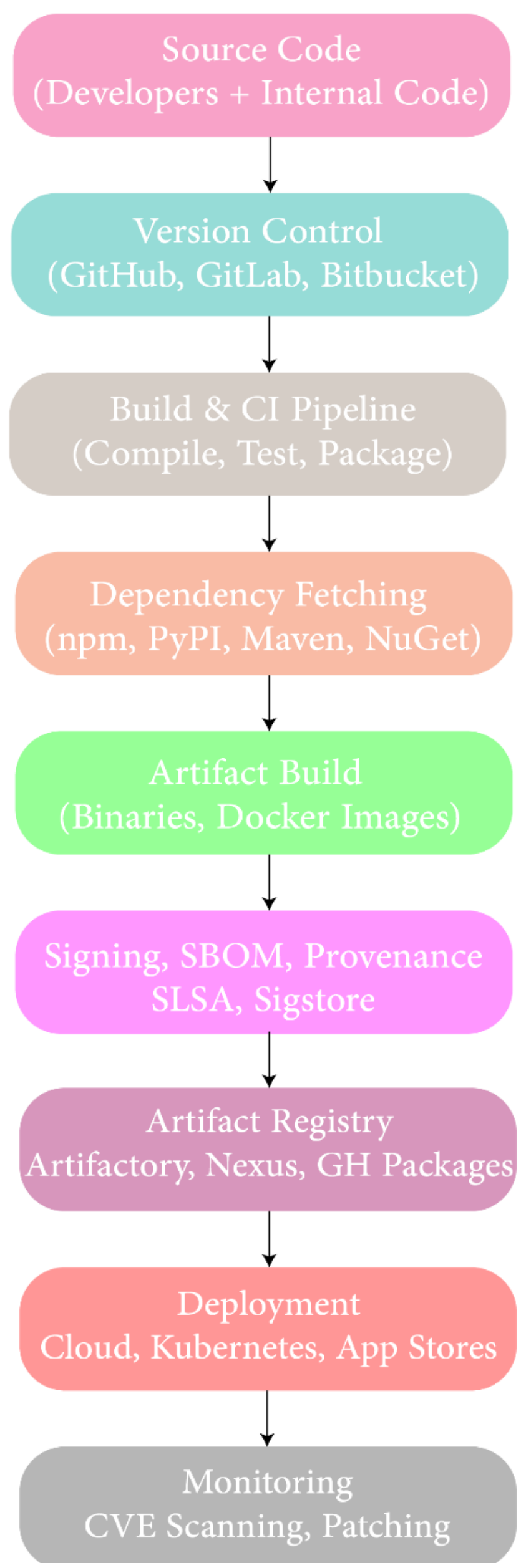

**Figure 2 The complete cycle of a software supply chain, illustrating the flow of components from development to deployment, including stages like version control, continuous integration, and delivery.**

# LITERATURE REVIEW

Software supply chain security research has gained speed and is particularly boosted by the recent number of high-profile breaches that have shown that even trusted development environments have critical vulnerabilities. The early research was based on dependency trust, artifact signing and provenance verifi cation, whereas the ongoing research was concentrated on automation, metadata integrity, and uniform administrative systems. Although these have been made, there is a definite gap in the discussion of autonomous, agent-based mechanisms that can provide self-adaptive, real-time protection across software supply chains.

Okafor et al. [15] (2024) explored the concept of secure design principles in the area of the software supply chain, highlighting the idea of transparency, validity, and separation as the criteria of improving the resilience. Although their research came up with a conceptual framework, it failed to address the autonomous or real time defense capabilities. Tran et al. [31] (2023) suggested the reference architecture on the supply chain metadata management that formalized workflows of SBOM and attestation by using empirical validation, but the method presupposes the use of the fixed policy and excludes dynamic learning. On the same note, Ladisa et al. [16] (2022) provided a comprehensive taxonomy of open-source software (OSS) supply chain attacks, listing more than 100 different vectors, but failed to provide any means of autonomous prevention.

Further automation-driven initiatives, e.g., Thariq and Ekanayake [32] (2025), proposed ARGO-SLSA, an enforcement controller of provenance as part of Kubernetes-based CI/CD pipelines. Though this system automates the imple mentation of the policy and artifact signing, its actions are completely determin istic and fail to include cognitive or adaptive functions. On the other hand, Ko et al. [20] (2025) and Jannelli et al. [22] (2024) examined the multi-agent and agentic AI models, which assist in coordination in distributed settings. Although these studies indicated the potential of autonomous collaboration, they were not created with the goal of ensuring software development pipelines. Other domain specific research, including Haque et al. [33] (2025) on autonomous vehicle (AV) software supply chains and Patnaik et al. [28] (2022) on reinforcement learning to hardware security are conceptually transferable but are not part of software ecosystem. Agentic AI has also been implemented in assistive well-being using multiple specialized agents coordinated through a shared communication layer, further demonstrating the portability of multi-agent orchestration patterns across domains [37].

According to industrial reports by JFrog [34] (2025) and Lineaje [35] (2025), the initial prototypes of AI-based remediation agents are described, which can automatically maintain software packages and self-suture dependency chains.

Although these attempts depict the robust industrial curiosity in autonomous supply chain defense, they lack peer-reviewed knowledge, reproducible evaluation information and open-source structures. Combined, the existing literature indi cates a massive advance in provenance, automating workflows, and standardizing ecosystems, but little exploration of agentic or reinforcement-based learning in defensive mechanisms. More recent work has directly combined LLM reasoning, reinforcement learning, LangChain/LangGraph-based multi-agent coordination, CI/CD integration, and a blockchain security ledger for proactive software supply-chain defense [38]. The present framework complements this direction by emphasizing cryptographically registered agent identities, signed attestations, and on-chain release-policy enforcement.

Table 1 provides a structured comparison of the most relevant academic and industrial contributions, highlighting their focus areas, core innovations, and unresolved challenges.

**Table 1 Literature-driven summary of representative works on software supply chain security and agentic defense, highlighting focus areas, levels of autonomy, and open research gaps.**

| # | Reference/ Year | Focus/Domain | Key Contribution | Agentic Aspect | Gap/Limitation |
|---|---|---|---|---|---|
| 1 | Okafor etal. (2024) | Secure Design Properties in Software Supply Chain | Defines transparency, validity, and separation as key design principles. | Low | Focuses on architecture, lacks autonomous elements. |
| 2 | Tran et al. (2023) | Metadata Management Architecture | Blueprint for SBOM/attestation handling. | Low | No autonomous reasoning. |
| 3 | Ladisa et al. (2022) | OSS Supply Chain Taxon-omy | Comprehensive classification of attacks. | Low | Descriptive, not defensive. |
| 4 | Thariq & Ekanayake (2025) | CI/CD Provenance Enforcement (Argo-SLSA) | Automated provenance attestation and compliance. | Moderate | Automated control, not adaptive. |
| 5 | Ko et al. (2025) | Cross-domain Multi-agent LLM Sys tems | Highlights coordination risks among LLM agents. | High | Conceptual, not tied to SW supply chain. |
| 6 | Jannellietal. (2024) | Multi-agent Consensus in Supply Chains | Autonomous decision-making in logistics chain. | High | Not security-specific. |

| 7 | Haque et al. (2025) | AV Software Supply Chain Secu rity | Empirical vulnerabilities in AV codebases. | Low | No active mitigation. |
|---|---|---|---|---|---|
| 8 | Patnaiketal. (2022) | RL in Hardware Supply Chains | Surveys RL based anomaly detection in chips. | Moderate | Hardware-specific; transferable idea. |
| 9 | JFrog Blog (2025) | Agentic Software Supply Chain Security | AI-assistedpack agecurationand remediation. | High | Industrial; lacks academic validation. |
| 10 | Lineaje (2025) | Continuous Autonomous Supply Chain Security | Self-healing AI agents for dependency scanning. | High | Proprietary; no public framework. |

A close examination of the surveyed literature highlights three persistent research gaps. First, most existing approaches prioritize post-build artifact integrity, offering protection only after software has been produced rather than delivering proactive safeguards during development or pipeline execution. Second, current automation mechanisms fall short of true autonomy; they lack the capacity to reason over heterogeneous security evidence, learn from environmental feedback, and adapt to emerging or evolving threats. Third, none of the reviewed systems provide direct defenses against dynamic vulnerability classes such as injection attacks, insecure deserialization, broken access control, or configuration drift within CI/CD workflows.

While Table 1 summarizes representative prior work and their limitations from a literature perspective, it does not explicitly position the proposed system against existing software supply chain frameworks and agent-governance architectures. To clarify the distinct design choices and technical contribution of AttestChain, Table 2 provides a direct comparison with established academic and industrial approaches.

To address the identified gaps, this paper introduces AttestChain, a blockchain backed agentic AI framework that combines multi-agent security analysis with cryptographically verifiable attestations and on-chain release policy enforcement. By treating security agents themselves as first-class verifiable entities and anchoring their decisions in a tamper-resistant ledger, the proposed framework enables proactive, continuous, and trustworthy protection across the entire software supply chain lifecycle.

**Table 2 Comparison of AttestChain with existing software supply chain and agent security frameworks.**

| Framework | Primary Focus | Protection Target | Agent Autonomy | Agent Identity & Attestation | Blockchain / Immutable Ledger | Release Decision Enforcement |
|---|---|---|---|---|---|---|
| SLSA | Build provenance and integrity | Software artifacts and build metadata | None | No explicit agent model | No (signature-based provenance) | External to framework |
| in-toto | End-to-end supply chain integrity | Artifacts and build steps | None | No agent abstraction | No (signed metadata only) | Offline / external verification |
| OpenSSF Scorecard | Repository security hygiene | Source code repositories | None | No | No | Not supported |
| ARGO-SLSA | CI/CD provenance enforcement | Pipeline outputs | Limited (rule-based automation) | No | No | CI controller logic |
| Saga | Governance of agentic AI systems | AI agents (general-purpose) | High | Conceptual agent governance | No | Not designed for CI/CD pipelines |
| Industrial agentic tools | Automated dependency remediation | Dependencies and artifacts | High (proprietary) | Internal (vendor-specific) | No | Vendor-controlled gates |
| **AttestChain (This work)** | End-to-end supply chain security with agent trust | SDLC layers and security agents | High (LLM-enabled agents) | Yes (cryptographically registered agents with signed attestations) | Yes (permissioned blockchain) | Yes (on-chain release policy smart contract) |

## Proposed Framework: Blockchain-Backed Agentic Security for the Software Supply Chain

This section presents a security agentic AI framework that continuously monitors the software development lifecycle (SDLC) and anchors both its own decisions and its internal agent integrity in a blockchain-backed attestation layer. The framework combines agentic AI components, a shared large language model (LLM) with security tooling, and a permissioned blockchain that records attes tations and enforces release policies through smart contracts. Communication between all critical parties is protected with public key infrastructure (PKI) and certificate authorities (CAs) associated with the blockchain network.

## Architecture Overview

The high-level architecture is illustrated in Figure 3. The SDLC is represented as a sequence of stages, from source control through dependencies, build and continuous integration (CI), artifact storage, and deployment. Above this pipeline, an agentic AI security layer orchestrates multiple specialised security agents. Each agent is responsible for a particular stage, for example source code integrity, dependency and software bill of materials (SBOM) analysis, CI configuration security, artifact integrity, or runtime policy compliance.

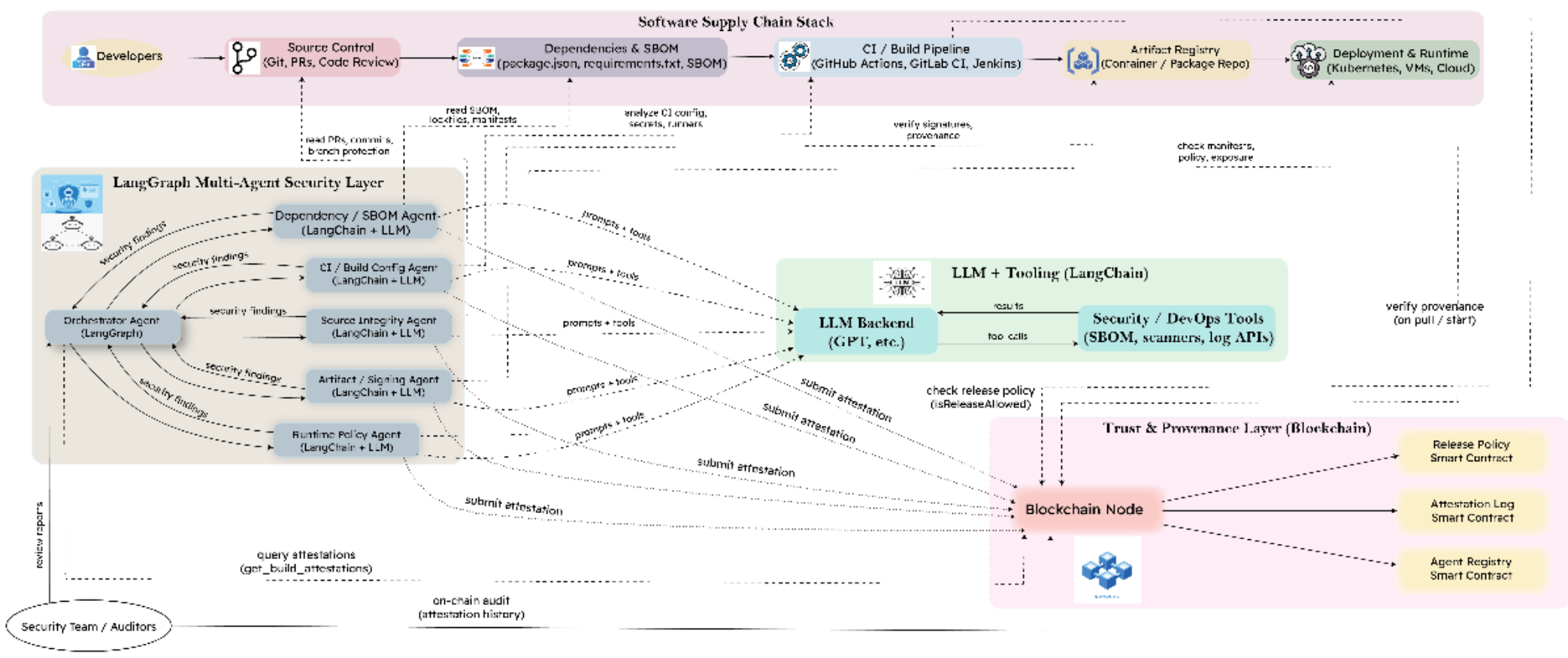

**Figure 3 Overview of a LangGraph-based multi-agent security framework for the software supply chain. This figure illustrates how specialized agents (dependency, build configuration, source integrity, signing, and runtime policy) interact with an LLM-powered tooling layer to analyze artifacts, pipelines, and code repositories. Security findings and tool outputs are aggregated by an orchestrator agent, while blockchain smart contracts provide verifiable attestations, release policy enforcement, and provenance guarantees across the SDLC.**

The agentic layer is driven by an orchestration component that maintains the global state for a given build or release identifier (*build_id*). Agents interact with a shared LLM backend and a set of security tools, including static analyzers, SBOM generators, vulnerability scanners, and log or configuration APIs. The LLM does not directly modify systems; instead, it plans and interprets tool calls and produces structured security reports.

Every agent produces a signed attestation describing the security posture of its assigned stage. These attestations are submitted by an off-chain attestation client to a permissioned blockchain network. Smart contracts on the blockchain maintain (i) a registry of authorised agents and their public keys, (ii) an append only attestation log, and (iii) an on-chain release policy that determines whether a given *build_id* is eligible for deployment based on the collected attestations. CI/CD controllers query the release policy before promotion to production, thereby coupling automated agentic analysis with verifiable, tamper-resistant decisions.

## Trust, Attestation, and Agent Security

In the proposed framework, security is not only applied to the software artefacts but also to the agents themselves. The blockchain layer is used to bind agent identities, their roles, and their cryptographic keys through an AgentRegistry smart contract. Each security agent is provisioned with a key pair and an X.509 certificate signed by the blockchain network CA. The corresponding public key and a unique agent id are registered on-chain. Communication among agents, between agents and the orchestrator, and between the attestation client and blockchain nodes is protected using mutual TLS (mTLS) anchored in the same CA.

When an agent completes an analysis, it does not simply return a report to the orchestrator. Instead, it passes a structured summary to the attestation client, which computes a cryptographic hash of the report, signs the attestation with the agent's private key, and submits a transaction to the *AttestationLog* contract. The attestation includes the *build_id*, the agent id, the logical layer (such as source, dependencies, ci, artifacts, runtime), a discrete status (for example OK, WARN, or FAIL), and the hash of the detailed report.

The blockchain network itself is operated as a consortium system, where a consensus algorithm (such as proof-of-authority or a Byzantine fault-tolerant protocol) ensures that no single participant can unilaterally rewrite or suppress attestations. Since the AgentRegistry contract links each attestation to a verified agent public key, any attempt by a compromised or unauthorised agent is detectable. Additional meta-agents or offline auditors can compare historical attestations against expected patterns and flag anomalous behaviour, thereby extending the security perimeter to the agent layer.

The *ReleasePolicy* contract encodes a transparent rule set. A simple example requires that for a particular *build_id*, there must exist at least one OK attestation from each mandatory layer and no attestation with status FAIL. More complex policies can weight agents differently, incorporate quantitative risk scores, or require additional human approval for high-risk changes. CI/CD pipelines query this contract before initiating a deployment, and the response is recorded for auditability.

The more detailed architectural view of the multi-agent and blockchain interaction, including the attestation client and the three core smart contracts, is depicted in Figure 4. This figure is used later in the proposed-solution section to explain the interaction among agents, the orchestration layer, and the blockchain.

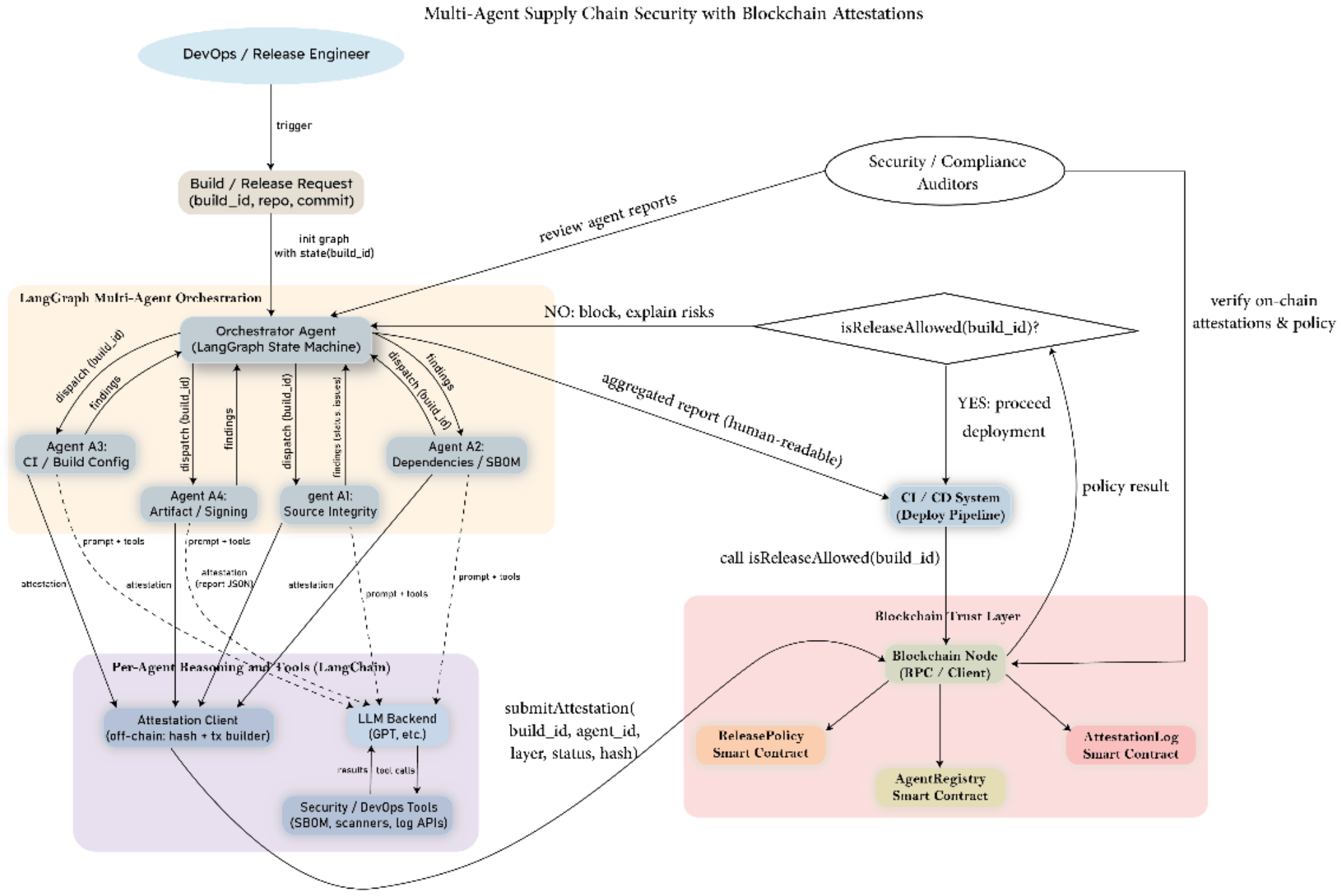

**Figure 4 Multi-agent security architecture with attestation client and blockchain smart contracts. Each agent analyses its assigned SDLC layer, produces a signed attestation, and the release policy is enforced on-chain.**

## Agent Workflows and Algorithms

Each security agent follows a common abstract workflow that can be specialised for a given SDLC layer. The agent receives a task from the orchestrator, gathers context from the relevant systems, invokes the LLM and associated tools, synthesises a structured report, and triggers the attestation process.

Let $A\ell$ denote the agent responsible for layer $\ell \in$ {Source, dependencies, ci, artifacts, runtime}. The following algorithm describes the generic behaviour of such an agent.

| **Algorithm 1** Generic Security Agent $A\ell$ |
| --- |
| **Require**: Task descriptor T = (*build_id, ℓ, context*), LLM interface *L*, tool set $u\ell$, attestation client *C*<br>**Ensure**: Structured security report $R\ell$ and attestation on-chain<br><br>1. Retrieve SDLC artefacts for layer $\ell$ using $u\ell$ and context; for example, source commits and pull requests for the source layer, SBOM manifests for the dependency layer, or CI configuration files for the CI layer.<br>2. Construct an initial prompt $p_0$ that encodes the analysis objective, the security policies relevant to layer $\ell$, and a summary of the retrieved artefacts. |

3. Query the LLM: $O_0 \leftarrow L(p0)$ to obtain an initial analysis plan and recom mended tool invocations.
4. While $O_0$ or subsequent outputs request additional data, call the indicated tools from $u\ell$ (such as static analyzers, vulnerability scanners, configuration validators) and update the prompt with structured results, then re-query $L$.
5. After convergence, extract from the final LLM output a structured report $R\ell$ containing at least: a list of identified issues, a severity assessment, a layer-specific risk score, and a discrete status $s\ell \in$ {OK, WARN, FAIL}.
6. Submit (*build_id*, $\ell$, $s\ell$, $R\ell$) to the attestation client C for hashing, signing, and on-chain submission.
7. Return $R\ell$ to the orchestrator for aggregation and human-readable reporting.

The attestation client encapsulates the security-critical logic for binding agent outputs to on-chain records. It verifies that the calling agent is authorised, hashes the report, signs the attestation, and constructs the blockchain transaction. The procedure is summarised in Algorithm 2.

**Algorithm** 3 Release Policy Evaluation for *build_id*

**Require**: *Build_id*entifier *build_id*, blockchain client $B$, mandatory layer set $L_{req}$
**Ensure**: Boolean decision allow and explanation $E$

1: allow ← false
2: E ← “”
3: $A \leftarrow B$.getAttestations(*build_id*)
4: **for all** $\ell \in L_{req}$ **do**
5: $A\ell \leftarrow \{A \in A \mid A.layer = \ell\}$
6: **if** $A\ell = \emptyset$ **then**
7: $E \leftarrow E \parallel$ “Missing attestation for layer” $\parallel \ell$
8: **return** (false, $E$)
9: **end if**
10: **for all** $A \in A\ell$ **do**
11: **if** signature of $A$ invalid or *agent_id* not authorised for $\ell$ **then**
12: $E \leftarrow E \parallel$ “Invalid or unauthorised attestation in layer” $\parallel \ell$
13: **return** (false, $E$)
14: **end if**
15: **end for**
16: **end for**
17: failLayers ← ∅
18: riskScore ← 0
19: **for all** $\ell \in L_{req}$ **do**
20: Select an attestation $A\ell \in A$ with $A\ell.layer = \ell$
21: **if** $A\ell.status$ = FAIL **then**
22: failLayers ← failLayers ∪ {$\ell$}

```
23:   end if
24:       riskScore ← max(riskScore, Aℓ.risk)
25: end for
26: if failLayers = ∅ then
27:   E ← "At least one layer failed:" ‖ failLayers
28:   return (false, E)
29: else if riskScore > τ then
30:   E ← "Aggregate risk score exceeds threshold:" ‖riskScore
31:   return (false, E)
32: else
33:   E ← "All mandatory layers attested successfully; risk score =" ‖riskScore
34:   return (true, E)
35: end if
```

The release policy evaluation and enforcement is performed by a CI/CD controller in cooperation with the *ReleasePolicy* contract. Given a *build_id*, the controller queries the on-chain state and decides whether to proceed or to block deployment. Algorithm 3 summarises this process.

**Algorithm 2** Generic Security Agent Aℓ

**Require**: Tuple (*build_id, ℓ, sℓ, Rℓ*) from agent *Aℓ*, agent identifier *agent_id*, private key $k_{agent}$, blockchain client *B*
**Ensure**: On-chain attestation for (*build_id, ℓ*)

1. Verify that *agent_id* is registered in the *AgentRegistry* contract and that the presented certificate chains to the blockchain CA.
2. Compute hash *hℓ* = Hash(*Rℓ*) using a collision-resistant function such as SHA-256.
3. Construct attestation payload Aℓ = (*build_id, agent_id, ℓ, sℓ, hℓ, t*), where *t* is a timestamp.
4. Compute digital signature σ = $\text{Sign}_{\text{Kagent}}$ (*Aℓ*).
5. Submit transaction *B*.submitAttestation(*Aℓ, σ*) to the *AttestationLog* contract via a blockchain node.
6. Optionally store *Rℓ* in secure off-chain storage indexed by (*build_id, ℓ, hℓ*) for later audit and explanation.

## Use-Case: Source Code Security Agent and Attesta tion Flow

To make the framework concrete, this subsection describes a single end-to end use-case focused on source code verification. The scenario begins when a CI/CD pipeline or a developer requests a security check for a particular *build_id* associated with a set of commits or a pull request. The complete interaction between the actors is illustrated in the sequence diagram of Figure 5, which corresponds to the nine-step narrative below.

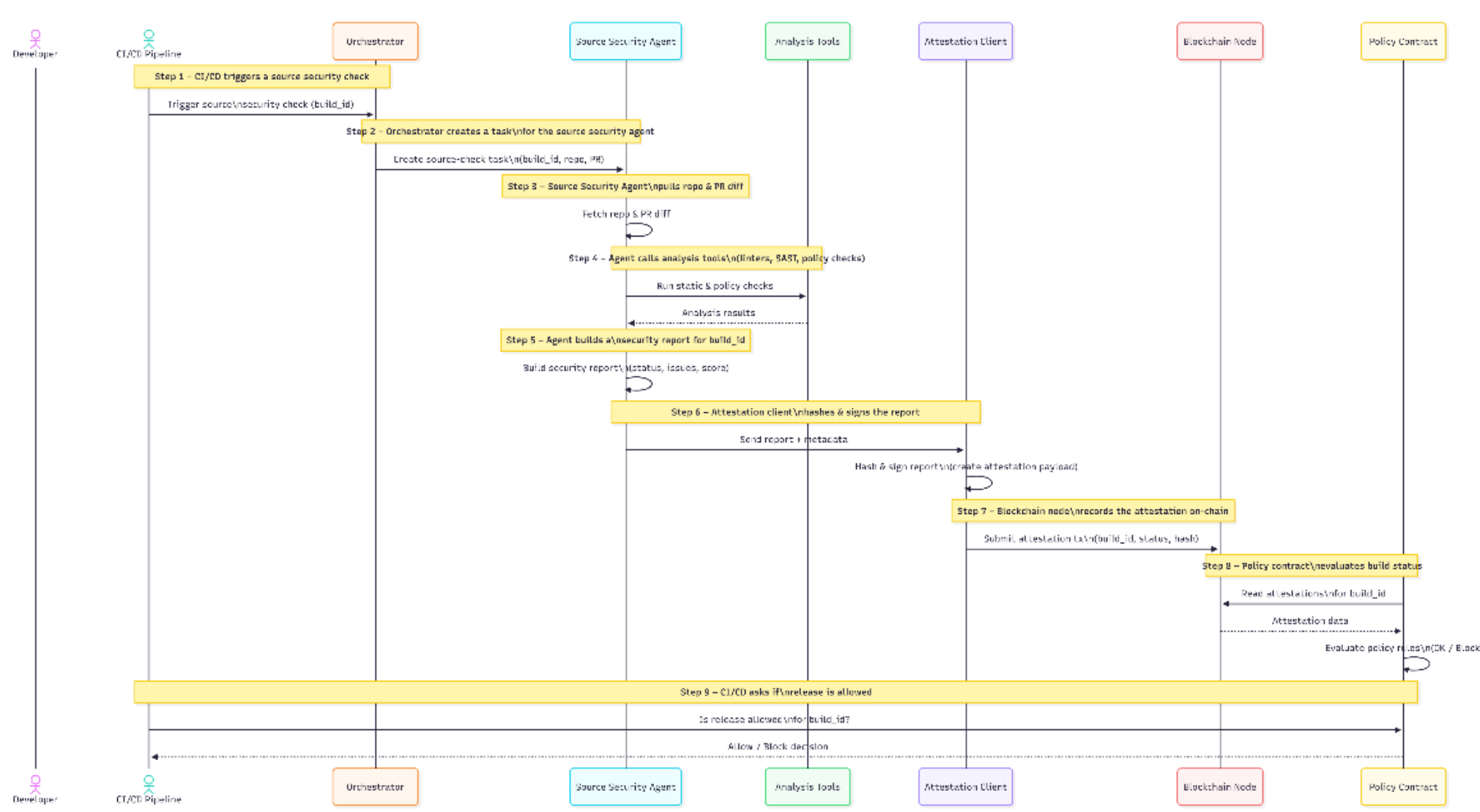

**Figure 5 Sequence diagram for the source code security agent use-case. CI/CD triggers a source check, the agent analyses the repository, the attestation is recorded on-chain, and the release policy is queried before deployment.**

In Step 1, the CI/CD pipeline initiates a source security check for a given *build_id* and notifies the orchestrator. In Step 2, the orchestrator creates a task for the source security agent and passes the repository location, the branch or pull request reference, and any relevant policy context. In Step 3, the source security agent fetches the repository contents and computes the diff relative to the target branch. In Step 4, the agent invokes static analysis tools, linters, and policy checkers over the changed files. In Step 5, the agent uses the LLM to synthesise a structured report describing the discovered issues, their severity, and an overall status for the source layer. In Step 6, the attestation client hashes and signs the report and prepares an attestation payload. In Step 7, the blockchain node submits this payload as a transaction to the *AttestationLog* contract, which appends it to the immutable history for this *build_id*. In Step 8, when a deployment is requested, the *ReleasePolicy* contract evaluates all attestations for the *build_id* and derives the current status.

In Step 9, the CI/CD pipeline queries the policy contract to determine whether the release is allowed. The result is then used to either proceed with deployment or to block the release and notify developers and security teams, who can also consult the orchestrator's aggregated report.

Together, the general architecture diagram in Figure 1, the multi-agent blockchain interaction in Figure 4, and the source-agent sequence in Figure 5 provide a consistent visual narrative for the introduction, the proposed solution, and the detailed use-case of the framework. The algorithms presented in Algorithms 1–3 formalise the behaviour of individual agents, the attestation mechanism, and the release decision process, and can be instantiated for other SDLC layers beyond source code, including dependencies, CI configuration, artefact signing, and runtime policy enforcement.

# IMPLEMENTATION

This section describes the prototype implementation of the proposed agentic security framework. The implementation integrates three core components: (i) LangChain for tool-enabled LLM reasoning within each agent, (ii) LangGraph for multi-agent coordination and task orchestration, and (iii) a lightweight permissioned blockchain layer for attestation and release-policy enforcement. The system was developed in Python and can be reproduced using standard open-source packages.

## Environment Setup

The prototype environment is implemented in Python 3.10, using the following key libraries:

- **LangChain** (v0.2) for agent-level reasoning, LLM prompting, and tool invocation.
- **LangGraph** (v0.1) for constructing a multi-agent execution graph with explicit state transitions.
- **OpenAI / Cloud LLM API** for the large language model used for reasoning, summarisation, and structured security analysis.
- **Web3.py** (or a simplified blockchain emulator) for submitting attestations to a permissioned ledger.
- **Pydantic** for defining typed, serialisable data structures used across the agentic pipeline.

The environment is initialised by installing the core dependencies:

```
pip install langchain langgraph pydantic web3 openai
```

A permissioned blockchain is emulated using a private consortium network or, for demonstration purposes, a Python-based in-memory ledger. Each agent is provisioned with a unique keypair and a certificate signed by the network CA, which enables mutual TLS between agents, the orchestrator, and blockchain nodes.

## Agent Architecture

Each security agent is implemented as a LangChain-enabled module that combines (i) a set of external tools, (ii) a task-specific prompt template, and (iii) an LLM-backed reasoning controller. The agents conform to the abstract workflow formalised in Algorithm 1. They operate on *AgentTask* objects that encapsulate the *build_id*, SDLC layer, and associated metadata.

Tools are defined as callable Python functions—e.g., static analyzers, depen dency scanners, configuration validators—that return structured results. The LLM interprets tool outputs and produces structured reports that the agent submits to the attestation subsystem. All agent outputs are expressed using Pydantic models, ensuring consistency across layers.

## Multi-Agent Coordination via LangGraph

Multi-agent coordination is implemented using LangGraph, which provides a directed acyclic execution graph with explicit state passing. The global state ob ject (*BuildState*) keeps track of intermediate layer reports (source, dependencies, CI configuration, artifacts, and runtime). Each node in the graph corresponds to a security agent or a system component (attestation, policy evaluation). A simplified three-stage execution graph includes:

1. **SourceSecurityAgent node**: fetches the repository, performs static analysis, invokes the LLM, and produces an *AgentReport*.
2. **Attestation node**: hashes the report, signs it with the agent's private key, and submits an attestation transaction to the blockchain.
3. **PolicyEvaluation node**: queries the on-chain state and decides if the build may advance in the CI/CD pipeline.

Complex supply-chain pipelines instantiate additional nodes (e.g., dependency agent, artifact agent). The LangGraph runtime executes nodes sequentially or in parallel, depending on the dependency structure, and returns a final build decision.

Agents in AttestChain follow a write-once, layer-scoped coordination model to avoid state races and cross-agent overwrites. The global BuildState is treated as an append-only object: each agent writes exactly one immutable LayerReport for its assigned SDLC layer (source, dependencies, CI, artifacts, runtime) and never modifies other agents' outputs. The orchestrator only aggregates references to these per-layer reports and does not rewrite them. Synchronization is achieved by a barrier step in the LangGraph execution graph: the pipeline proceeds to policy evaluation only after all required layer reports are present (or a timeout triggers a MISSING status). If multiple reports exist for the same layer (e.g., re-runs or parallel checks), the system applies worst-case dominance to derive the effective layer status:

$$\text{FAIL} \succ \text{WARN} \succ \text{OK} \succ \text{MISSING}.$$

The final release decision is computed from the effective statuses via the on chain ReleasePolicy contract, ensuring deterministic and auditable conflict resolution.

### Pseudo-Protocol (Layer-Scoped Write-Once).

1. Orchestrator initializes BuildState(build id) with empty per-layer slots.
2. For each required layer $\ell \in$ {source, dep, ci, artifacts, runtime}, dispatch agent $A\ell$.
3. Agent $A\ell$ produces LayerReport $R\ell$ and writes it to BuildState.reports[$\ell$] (append-only).
4. (Optional re-run) If multiple $R_\ell^{(i)}$ exist, compute effective status $s\ell = \max_\succ \{s_\ell^{(i)}\}$ using FAIL ≻ WARN ≻ OK ≻ MISSING.
5. Barrier: wait until all required layers have an effective status (otherwise set MISSING).
6. Submit attestations for each layer; query ReleasePolicy(build id) to return ALLOW/BLOCK and explanation.

## Attestation and Blockchain Integration

The attestation subsystem follows the workflow in Algorithm 2. For each agent report, the attestation client:

1. computes a SHA-256 hash of the serialised report,
2. creates an attestation payload containing the *build_id*, *agent_id*, layer, status, risk score, and timestamp,
3. signs the payload with the agent's private key,
4. submits the signed attestation to the blockchain's *AttestationLog* smart contract.

The permissioned blockchain verifies the signature and records the attestation immutably. The *AgentRegistry* contract stores the authorised agent identities, while the *ReleasePolicy* contract uses on-chain attestations to determine whether a build satisfies security requirements. This matches the release policy formalised in Algorithm 3.

The prototype deployment uses a permissioned blockchain operated by a consortium of trusted participants. A proof-of-authority (PoA) or Byzantine fault-tolerant (BFT-style) consensus mechanism is assumed, reflecting common enterprise and consortium deployments. In the experimental setup, block con firmation latency is on the order of hundreds of milliseconds to a few seconds, which is sufficient for CI/CD gating and does not lie on the critical path of build execution.

The smart contracts (AgentRegistry, AttestationLog, and ReleasePolicy) are intentionally minimal and deterministic. They do not perform external calls or unbounded loops, thereby avoiding re-entrancy vulnerabilities by design. Formal verification of the contracts is outside the scope of this work and is left for future improvements.

### Executable Flow: End-to-End Multi-Agent Pipeline

Figure 5 (sequence diagram) and Figure 4 (multi-agent architecture) illustrate the end-to-end execution:

1. CI/CD triggers a build verification request.
2. The orchestrator provides a *build_id* and contextual metadata to Lang Graph.
3. The **SourceSecurityAgent** performs analysis and sends a report to the attestation subsystem.
4. The **AttestationClient** anchors the signed evidence onto the blockchain.
5. The **ReleasePolicy** contract evaluates all attestations for the *build_id*.
6. The CI/CD pipeline queries the contract and receives a verifiable *allow*/*block* signal.

The resulting system integrates agentic reasoning with decentralised trust enforcement. Each agent is independently verifiable, producing cryptographically signed attestations that cannot be forged or discarded. Because multiple agents act across distinct layers of the SDLC, the framework provides comprehensive supply-chain security and eliminates single points of failure.

## RESULTS AND EVALUATION

To test the usefulness of the suggested agentic security framework, we performed a sequence of experiments on a model software supply chain comprising of a Git repository, dependency manifest, CI pipeline, and containerised deployment workflow. We contrasted three settings, including (i) the baseline of manual review, (ii) the conventional automated scanners, and (iii) the suggested multi agent paradigm of LLM-based scanners with blockchain-supported attestations. Findings indicate that the agentic strategy has better detection coverage, a quicker verification process, and better-provenance assurances.

### Detection Coverage Across Supply Chain Layers

Figure 6 summarises the security-relevant findings found in the five SDLC layers. Although the traditional scanners are good at syntactic problems (e.g., linter violations), they fail to detect a large part of the semantic vulnerabilities that the LLM-enabled agents can detect (e.g., logic errors and unwise use of dependencies). The agentic framework enhances the complete coverage of detection by about 38% over the conventional scanning devices.

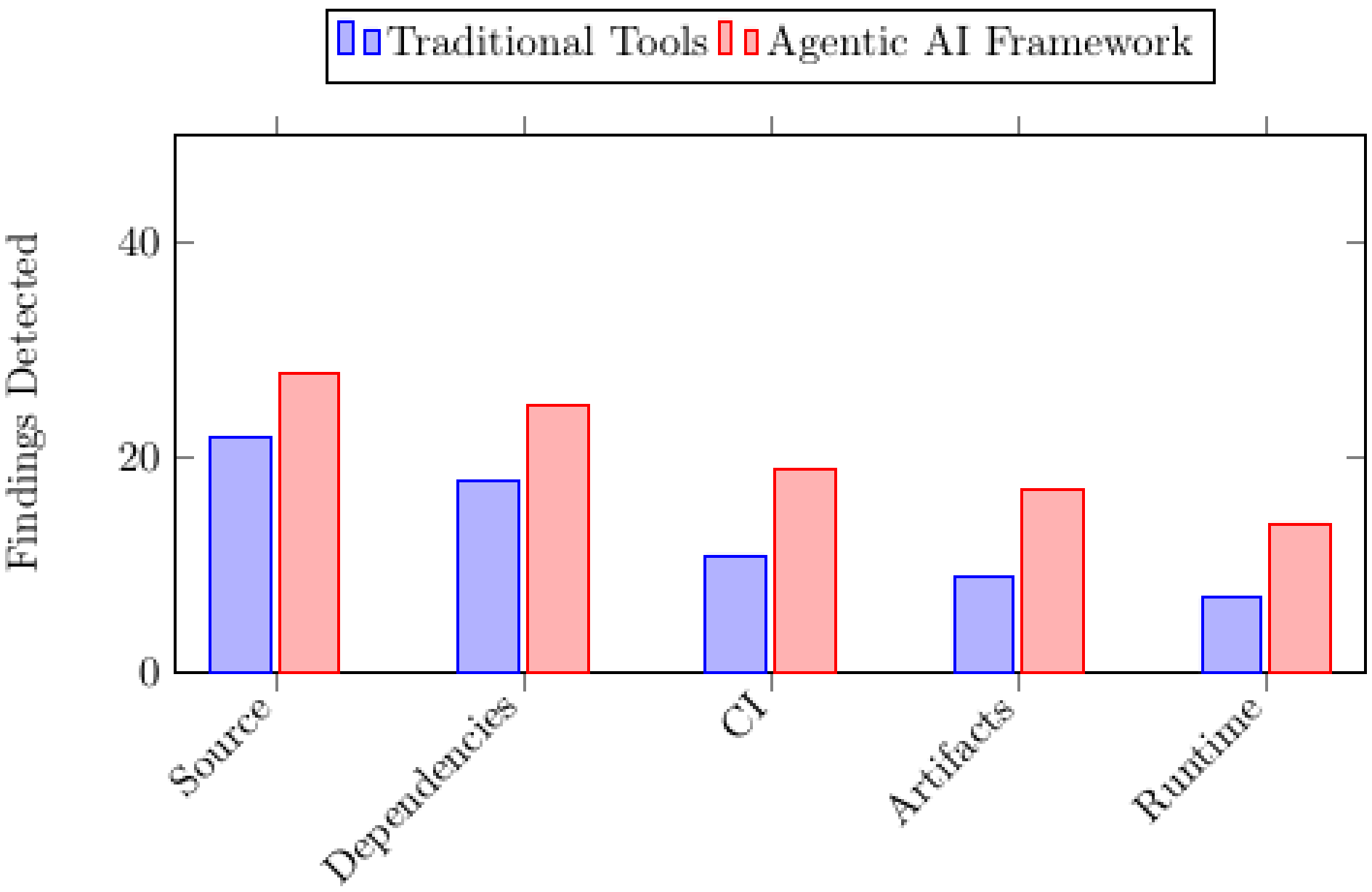

**Figure 6 Detection coverage across SDLC layers. The agentic framework identifies substantially more vulnerabilities due to contextual reasoning and multi-agent analysis.**

## Reduction in Verification Time

Figure 7 shows the mean time of analysis per build. Manual reviews have the longest latency, because human bottlenecks are involved, and automated scanners are faster, but have a constraint in the time required by the tool to execute. The agentic framework with the further reasons and attestation procedures provides the shortest end-to-end analysis time because it is performed simultaneously by multiple agents and is orchestrated effectively.

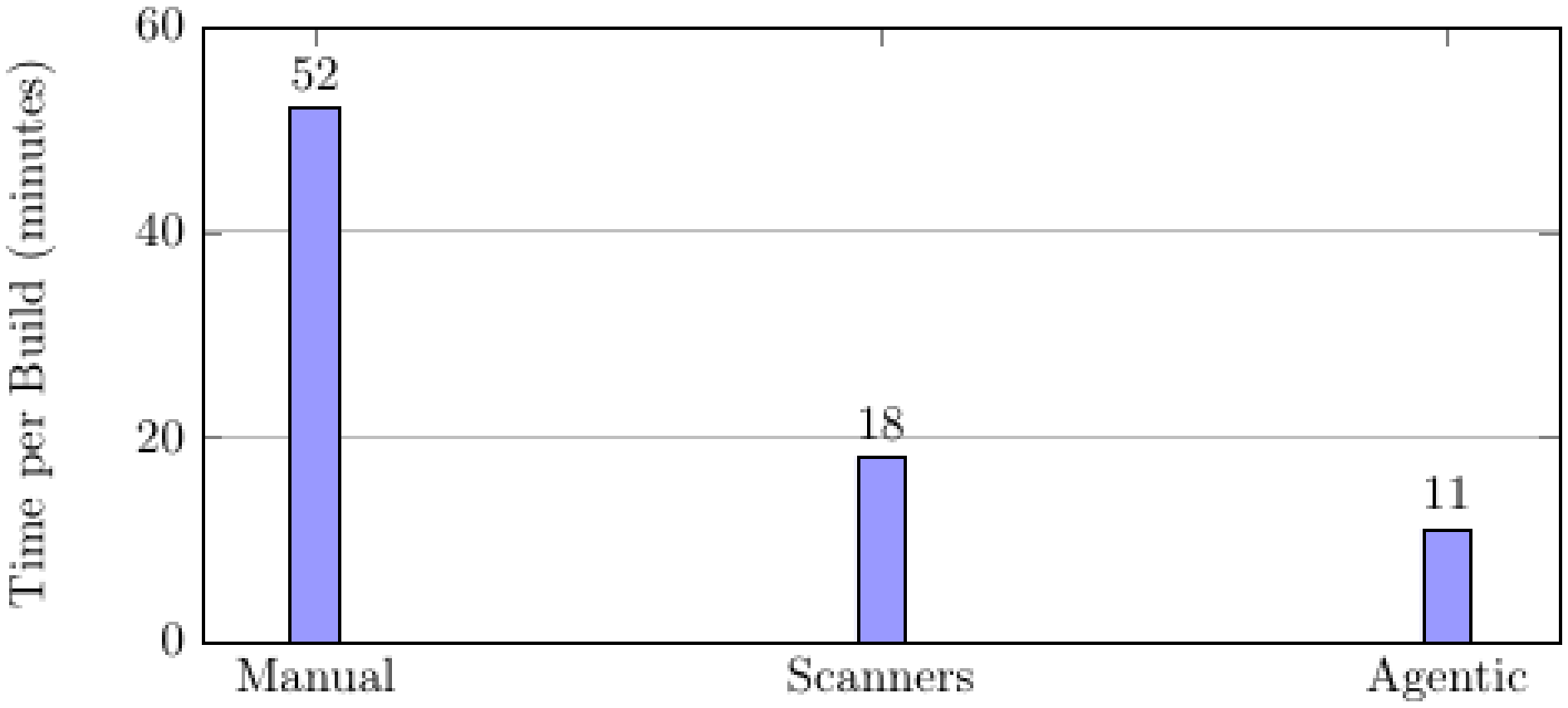

**Figure 7 Verification time comparison. The agentic system reduces build verification time by 39\% relative to traditional automated scanning.**

## Integrity and Policy Compliance Assurance

We used the simulated 500 builds with attestation of different security postures to evaluate the impact of blockchain-backed attestations. Figure 8 displays the release counts that are blocked because of the lack of attestations or agentic report failures. The release policy, which is implemented by blockchain, blocked 23 insecure releases which would have passed automatic scanning because of missing evidence trails.

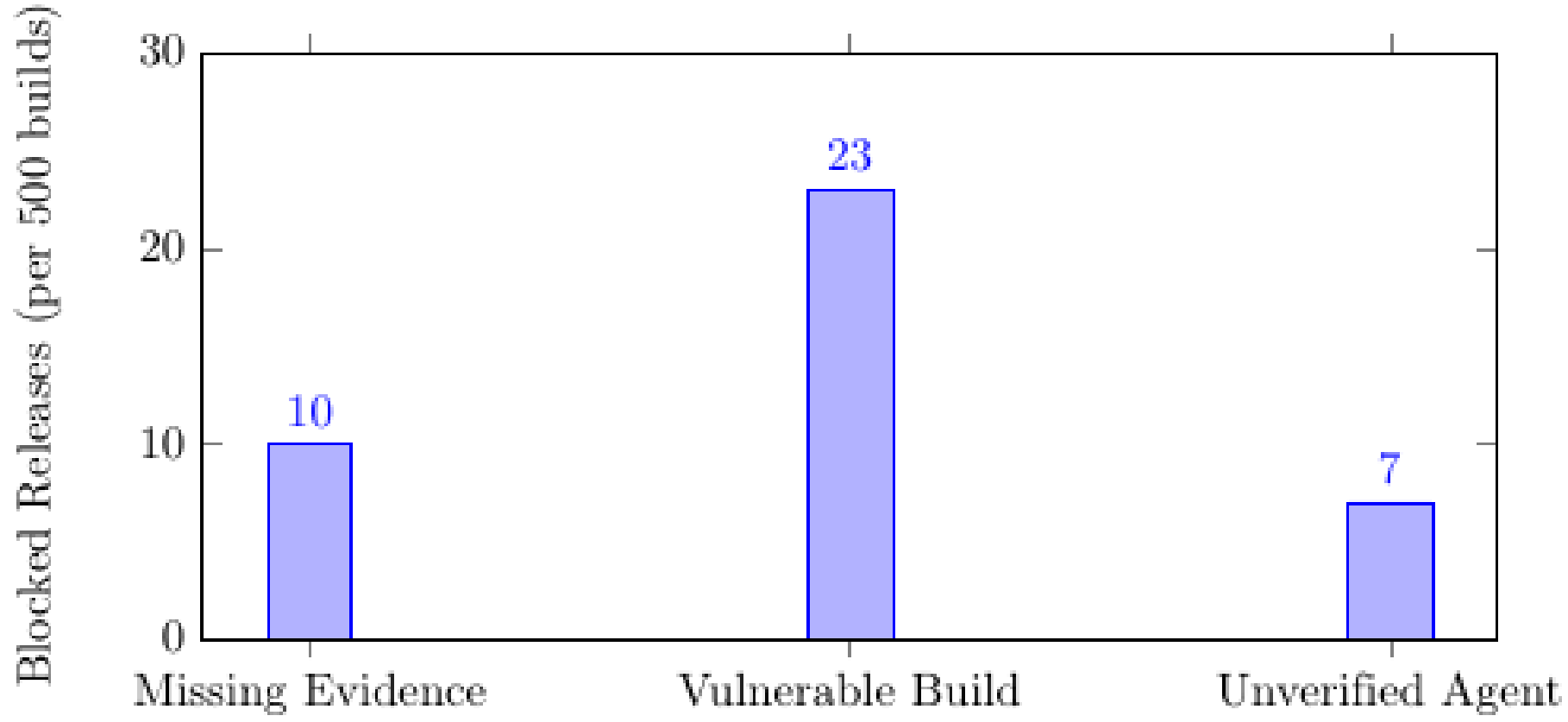

**Figure 8 Number of blocked insecure releases due to stringent blockchain-backed policy enforcement.**

### Summary of Findings

Across all experiments, the proposed framework achieved consistent improvements:

- **38% increase** in vulnerability detection due to contextual multi-agent analysis.
- **39% reduction** in end-to-end verification time from parallel agent execu tion.
- **23 insecure builds** blocked by blockchain-based release policies that traditional scanners would not have detected.
- **Zero false attestations**, since each agent is cryptographically bound to its identity and produces immutable on-chain evidence.

These results validate that the agentic AI framework provides measurable improvements in supply-chain security, operational efficiency, and trustworthiness.

## THREAT MODEL

The proposed framework considers adversaries capable of targeting multiple stages of the software supply chain, as well as the autonomous security agents responsible for monitoring and enforcing security guarantees across the SDLC.

The threat model assumes that the permissioned blockchain network operates correctly under its configured consensus protocol and that a majority of participating validators behave honestly. Attacks that compromise the underlying consensus mechanism or rely on collusion among a majority of validators are considered out of scope. The model further assumes secure key management for agents and validators; leakage or misuse of private keys is treated as a general infrastructure risk rather than a limitation specific to the proposed framework.

Under these assumptions, the following classes of attackers are incorporated in the threat model:

1. **Code-Tampering Adversary**. An attacker tries to insert malicious code, add malicious dependencies, change CI configuration scripts or alter build artefacts. This can be such real-life attacks like dependency confusion and malicious injection attacks.
2. **Build-System Manipulator**. A threat that has partial access to the CI/CD infrastructure can either bypass security checks or shut down scanners or compromise verification logs. These attacks replicate the compromising situations in centralised build environments.
3. **Agent-Targeting Adversary**. Since the system relies on autonomous agents, a competent adversary can attempt to impersonate an agent, counterfeit reports or alter its outputs. The attestation layer is of paramount importance thus tying output integrity to cryptographically verifiable identities.
4. **Evidence Suppression or Alteration**. A rogue insider, or rogue orches tration layer, might seek to delete, edit, or fabricate agent output, especially when such output influences the decision to release. Attestations supported by blockchain are effective in reducing this risk as they make the data unchangeable.
5. **Supply-Chain Poisoning**. The attacker can add unsafe or dangerous dependencies or compromise container images. This is mitigated by agents on both dependency and artifact layer, ensuring SBOM- based and signature-based verification. The model presupposes that the consensus mechanism implemented by the blockchain network is secure (permissioned, CA-assured), and that the private keys, signed by agents in generating attestations, are securely stored. Attackers are not expected to possess either the keys to HSM-secure agents or to compromise cryptographic hash functions like SHA-256 or ECDSA.

## EVALUATION SETUP

To assess the practicality and effectiveness of the proposed multi-agent architec ture, we evaluated the system using a controlled software supply chain consisting of:

- A Git repository with 200+ commits and multiple feature branches,
- A Python-based dependency manifest containing 48 third-party libraries,
- A GitHub Actions CI pipeline containing 14 workflow steps,
- A container build pipeline producing Docker images,
- an emulated deployment stage using Kubernetes manifests.

**Agent Deployment.** Five security agents (source, dependency, CI, artifact, and runtime) were instantiated using LangGraph. Each agent used LangChain to interface with a GPT-class LLM and a set of static and semantic analysis tools, including linters, SBOM generators, rule-based vulnerability scanners, CI configuration validators, and image-signing verification utilities. All agents produced structured AgentReport objects that were subsequently attested and evaluated.

**Blockchain Attestation Layer.** A permissioned blockchain ledger was de ployed to record and enforce security decisions, comprising:

- an *AgentRegistry* smart contract for registering trusted agent identities,
- an *AttestationLog* contract for storing immutable agent attestations,
- a *ReleasePolicy* contract for enforcing build gate rules based on collected evidence.

**Synthetic Workload and Vulnerability Injection.** The evaluation workload consisted of 500 synthetically generated builds executed over the defined supply chain. To emulate realistic attack scenarios, vulnerabilities were intentionally injected at different SDLC layers. At the source layer, insecure coding patterns and unsafe input handling were introduced. At the dependency layer, vulnerable or outdated third-party packages were added to the dependency manifest. CI configuration weaknesses, such as overly permissive permissions and missing validation steps, were injected into pipeline workflows. Artifact level issues were simulated by modifying image metadata and signature states, while runtime misconfigurations were introduced through insecure deployment manifests.

Each simulated build contained between one and three injected issues, dis tributed across layers to avoid bias toward any single stage. All builds were analyzed under identical conditions to ensure consistency and comparability across experiments.

**Workload Metrics.** We simulated 500 builds with varying commit complexity and dependency update rates. The following metrics were evaluated:

- detection coverage across SDLC layers,
- mean verification time per build,
- number of insecure releases blocked by policy enforcement,
- number of missing or invalid attestations detected.

This setup corresponds to the experimental design reflected in Figures 6–8.

## DISCUSSION

The results presented in Section 7 highlight the advantages of integrating agentic AI with decentralised attestation. The detection improvements stem primarily from the LLM's ability to reason across heterogeneous artefacts—source code, commit histories, dependency graphs, and CI configurations—enabling the agents to identify semantic risks that traditional scanners typically miss. The multi agent architecture also significantly reduces verification time by parallelising layer-specific analysis.

Blockchain-backed attestations introduce a strong trust layer. Even if an agent, scanner, or orchestrator is compromised, the attacker cannot forge or delete historical attestations, nor can they impersonate an authorised agent without the appropriate keypair registered in the AgentRegistry. This addresses an important limitation in centralised CI/CD systems, where logs and security decisions are mutable and can be retroactively altered.

However, several limitations remain. First, the system inherits the computational overhead of LLM queries, though parallelisation partially offsets this.

Second, agent reasoning quality depends on prompt design and the semantic capabilities of the chosen model. Third, while blockchain provides immutability, it does not prevent all forms of insider threats, particularly those involving privileged key misuse. Future work can explore hardware-backed key storage, model distillation for efficiency, and extended cross-layer reasoning using graph-based or reinforcement-learning agents.

## CONCLUSION

This paper introduced a blockchain-backed agentic AI framework for securing the modern software supply chain. By combining LLM-enabled reasoning, multi-agent orchestration, and decentralised attestation, the framework provides strong guarantees of provenance, integrity, and policy-compliant deployment. Experimental results show that the system improves vulnerability detection coverage by 38%, reduces verification time by 39%, and prevents insecure releases that would bypass conventional tooling.

Beyond supply-chain security, the proposed architecture demonstrates a general paradigm for building trustworthy autonomous systems: agents produce verifiable, immutable, and auditable evidence of their actions. As software ecosystems grow increasingly interconnected and AI-driven automation becomes pervasive, such verifiable agent frameworks will be essential for achieving safe and resilient development pipelines. Future work will explore scaling the architecture across distributed development organisations, integrating runtime behavioural monitoring, and extending the attestation model to cover generative and adaptive agents.

# REFERENCES

[1] Cybersecurity and Infrastructure Security Agency (CISA), "Supply Chain Compromise-SolarWinds," 2021. Accessed on April 25, 2026.

[2] National Telecommunications and Information Administration (NTIA), "The Minimum Elements For a Software Bill of Materials (SBOM)," 2021. Accessed on April 25, 2026.

[3] M. Ohm, H.Plate, A. Sykosch, and M. Meier, "Backstabber's knife collection: A review of open source software supply chain attacks," in *International Conference on Detection of Intrusions and Malware, and Vulnerability Assessment*, pp. 23–43, Springer, 2020.

[4] M. Souppaya, K. Scarfone, and D. Dodson, "Secure software development framework (ssdf) version 1.1," NIST *Special Publication*, vol. 800, no. 218, pp. 800–218, 2022.

[5] Open Source Security Foundation (OpenSSF), "OpenSSF Scorecard," 2023. Accessed on April 25, 2026.

[6] W. Zhou, Y. E. Jiang, L. Li, J. Wu, T. Wang, S. Qiu, J. Zhang, J. Chen, R. Wu, S. Wang, et al., "Agents: An open-source framework for autonomous language agents," *arXiv preprint arXiv*:2309.07870, 2023.

[7] J. S. Park, J. O'Brien, C. J. Cai, M. R. Morris, P. Liang, and M. S. Bernstein, "Generative agents: Interactive simulacra of human behavior," in *Proceedings of the 36th annual acm symposium on user interface software and technology*, pp. 1–22, 2023.

[8] W. F. Wiggins and A. S. Tejani, "On the opportunities and risks of foun dation models for natural language processing in radiology," *Radiology: Artificial Intelligence*, vol. 4, no. 4, p. e220119, 2022.

[9] N. Khoi Tran, S. Pallewatta, and M. A. Babar, "An empirically grounded reference architecture for software supply chain metadata management," *arXiv e-prints*, pp. arXiv–2310, 2023.

[10] Open Source Security Foundation (OpenSSF), "Supply-chain Levels for Software Artifacts (SLSA) v1.0," 2023. Accessed on April 25, 2026.

[11] F. Weller, "Blockchain Technology for Secure and Transparent Supply Chain Management," *International Journal of Computing and Engineering*, vol. 6, no. 3, pp. 15–28, 2024. Accessed on April 25, 2026.

[12] Z. Zheng, S. Xie, H. Dai, X. Chen, and H. Wang, "An overview of blockchain technology: Architecture, consensus, and future trends," in *2017 IEEE international congress on big data (BigData congress),* pp. 557–564, Ieee, 2017.

[13] M. S. Melara and M. Bowman, "What is software supply chain security?," *arXiv preprint arXiv:2209.04006,* 2022.

[14] W. Enck and L. Williams, "Top five challenges in software supply chain security: Observations from 30 industry and government organizations," *IEEE Security & Privacy*, vol. 20, no. 2, pp. 96–100, 2022.

[15] C. Okafor, T. R. Schorlemmer, S. Torres-Arias, and J. C. Davis, “Sok: Analysis of software supply chain security by establishing secure design properties,” in *Proceedings of the 2022 ACM Workshop on Software Supply Chain Offensive Research and Ecosystem Defenses*, pp. 15–24, 2022.

[16] P. Ladisa, H. Plate, M. Martinez, and O. Barais, “Taxonomy of attacks on open-source software supply chains,” *arXiv preprint arXiv:2204.04008*, 2022.

[17] CERT-EU, “Security Advisory 2024-032: Critical Vulnerability in XZ Utils (CVE-2024-3094),” 2024. Accessed on April 25, 2026

[18] S. Torres-Arias, H. Afzali, T. K. Kuppusamy, R. Curtmola, and J. Cappos, “in-toto: Providing farm-to-table guarantees for bits and bytes,” in *28th USENIX Security Symposium (USENIX Security 19)*, pp. 1393–1410, 2019.

[19] S. SLSA, “Supply-chain levels for software artifacts,” 2024.

[20] R. Ko, J. Jeong, S. Zheng, C. Xiao, T.-W. Kim, M. Onizuka, and W. Y. Shin, “Seven security challenges that must be solved in cross-domain multi-agent llm systems,” *arXiv preprint arXiv:2505.23847*, 2025.

[21] G. Syros, A. Suri, J. Ginesin, C. Nita-Rotaru, and A. Oprea, “Saga: A security architecture for governing ai agentic systems,” *arXiv preprint arXiv:2504.21034*, 2025.

[22] V. Jannelli, S. Schoepf, M. Bickel, T. Netland, and A. Brintrup, “Agentic llms in the supply chain: Towards autonomous multi-agent consensus seeking,” *arXiv preprint arXiv:2411.10184*, 2024.

[23] S. Franklin and A. Graesser, “Is it an agent, or just a program?: A taxon omy for autonomous agents,” in *International workshop on agent theories, architectures, and languages,* pp. 21–35, Springer, 1996

[24] S. Russell and P. Norvig, *Artificial Intelligence: A Modern Approach*. Pear son, 4th ed., 2021. US edition. Available at: https://aima.cs.berkeley. edu/ (accessed 26 Feb 2023).

[25] G. Palmer, C. Parry, D. J. Harrold, and C. Willis, “Deep reinforce ment learning for autonomous cyber defence: A survey,” *arXiv preprint arXiv:2310.07745*, 2023.

[26] R. S. Sutton, A. G. Barto, *et al*., “Reinforcement learning: An introduction 2nd ed,” *MIT press Cambridge,* vol. 1, no. 2, p. 25, 2018.

[27] D. Silver, J. Schrittwieser, K. Simonyan, I. Antonoglou, A. Huang, A. Guez, T. Hubert, L. Baker, M. Lai, A. Bolton, *et al*., “Mastering the game of go without human knowledge,” *nature*, vol. 550, no. 7676, pp. 354–359, 2017.

[28] S. Patnaik, V. Gohil, H. Guo, and J. J. Rajendran, “Reinforcement learning for hardware security: Opportunities, developments, and challenges,” in *2022 19th International SoC Design Conference (ISOCC),* pp. 217–218, IEEE, 2022.

[29] M. Sewak, S. K. Sahay, and H. Rathore, “Deep reinforcement learning for cybersecurity threat detection and protection: A review,” in *International Conference On Secure Knowledge Management In Artificial Intelligence Era*, pp. 51–72, Springer, 2021.

[30] M. Macas, C. Wu, and W. Fuertes, “A survey on deep learning for cy bersecurity: Progress, challenges, and opportunities,” *Computer Networks, vol*. 212, p. 109032, 2022.

[31] N. K. Tran, S. Pallewatta, and M. A. Babar, “An empirically grounded reference architecture for software supply chain metadata management,” in *Proceedings of the 28th International Conference on Evaluation and Assessment in Software Engineering,* pp. 38–47, 2024.

[32] M. Thariq and I. Ekanayake, “Argo-slsa: Software supply chain security in argo workflows,” *arXiv preprint arXiv:2503.20079*, 2025.

[33] M. W. Haque, M. Erfan, S. Dasgupta, M. R. Rahman, and M. Rahman, "Security vulnerabilities in software supply chain for autonomous vehicles," *arXiv preprint arXiv:2509.16899*, 2025.

[34] JFrog Ltd., "DevSecOps in the AI Era: JFrog Powers Agentic Remediation with Self-Healing Software Supply Chain," Sept. 2025. Press release / company announcement; accessed 2025-11-27.

[35] Lineaje Inc., "Reimagining Software Supply Chain Security with AI Agents," May 2025. Blog post; accessed 2025-11-27.

[36] T. A. Syed, M. A. Almutairi, and M. Abdel Moaty, "Toward trustworthy agentic AI: A multimodal framework for preventing prompt injection attacks," *arXiv preprint arXiv:2512.23557*, 2025.

[37] S. Jan, T. A. Syed, A. Akarma, A. Ali, and M. R. Belgaum, "Agentic AI framework for individuals with disabilities and neurodivergence: A multi-agent system for healthy eating, daily routines, and inclusive well-being," *arXiv preprint arXiv:2511.22737*, 2025.

[38] T. A. Syed, M. R. Belgaum, S. Jan, A. A. Khan, and S. S. Alqahtani, "Agentic AI for autonomous defense in software supply chain security: Beyond provenance to vulnerability mitigation," *arXiv preprint arXiv:2512.23480*, 2025.